\documentclass[12pt]{article}
\usepackage{graphicx}
\title{What causes a neuron to spike?}
\author{Blaise Ag\"uera~y~Arcas$^1$ and Adrienne L. Fairhall$^2$\\
$^1$Rare Books Library and $^2$Department of Molecular Biology,\\
Princeton University, Princeton, New Jersey 08544 \\
{\it blaisea@princeton.edu fairhall@princeton.edu} }

\begin{document}
\maketitle

\begin{abstract}

The computation performed by a neuron can be formulated as a
combination of dimensional reduction in stimulus space and the
nonlinearity inherent in a spiking output.  White noise stimulus
and reverse correlation (the spike-triggered average and
spike-triggered covariance) are often used in experimental
neuroscience to `ask' neurons which dimensions in stimulus space
they are sensitive to, and to characterize the nonlinearity of the
response.  In this paper, we apply reverse correlation to the
simplest model neuron with temporal dynamics---the leaky
integrate-and-fire model---and find that even for this simple case
standard techniques do not recover the known neural computation.
To overcome this, we develop novel reverse correlation techniques
by selectively analyzing only `isolated' spikes, and taking
explicit account of the extended silences that precede these
isolated spikes.  We discuss the implications of our methods to
the characterization of neural adaptation.  Although these methods
are developed in the context of the leaky integrate-and-fire
model, our findings are relevant for the analysis of spike trains
from real neurons.
\end{abstract}

\section{Introduction}
\label{s:introduction}

There are two distinct approaches to characterizing a neuron based
on spike trains recorded during stimulus presentation. One can
take the view of a downstream neuron, and ask how the spike train
should be decoded to reconstruct some aspect of the stimulus.
Alternatively, one can take the neuron's own view, and try to
determine how the stimulus is {\it en}coded, i.e. which aspects of
the stimulus trigger spikes.  This paper is concerned with the
latter problem: the identification of relevant stimulus features
from neural data.

The earliest attempts at neural characterization, including
classic electrophysiological experiments like those of Adrian
\cite{adrian}, presented the neural system with simple, highly
stereotyped stimuli with one or at most a few free parameters;
this makes it relatively straightforward to map the input/output
relation, or tuning curve, in terms of these parameters.  While
this approach has been invaluable, and is still often used to
advantage, it has the shortcoming of requiring a strong assumption
about what the neuron is `looking for'.  As this assumption is
relaxed by probing the system using stimuli with a larger number
of parameters, the length of the experiment needed to probe neural
response grows combinatorially.

In some cases, white noise can be used as an alternative stimulus
which naturally contains `every' signal \cite{marmarelis}. The
stimulus feature best correlated with spiking is then the spike
triggered average (STA), i.e. the average stimulus history
preceding a spike \cite{deBoer,segundo}.  Such reverse correlation
methods have been further developed and applied to characterize
the motion-sensitive identified neuron H1 in the fly visual system
\cite{bialek88}.  By extending reverse correlation to second
order, one can identify multiple directions in stimulus space
relevant to the neuron's spiking `decision'
\cite{naama-info-method,bill&robinprep}.  These techniques allow
the experimenter to probe neural response stochastically, using
minimal assumptions about the nature of the relevant stimulus.

In this paper, we will examine in detail the relation between the
stimulus features extracted through white noise analysis and the
computation actually performed by the neuron.  In the usual
interpretation of white noise analysis, the features extracted
using reverse correlation are considered to be linear filters
which, when convolved with the stimulus, provide the relevant
inputs to a nonlinear spiking decision function.  While the
features recovered by reverse correlation are optimal for stimulus
reconstruction \cite{spikes}, we will show that they are distinct
from the filter or filters relevant to the neuron's spiking
decision. In particular, reconstruction filters necessarily depend
on the stimulus ensemble, while the filters relevant for
spiking---at least, for simple neurons like leaky
integrate-and-fire---are properties of the model alone.

The main difficulty in interpreting the spike triggered average is
that spikes interact, in the sense that the occurrence of a spike
affects the probability of the future occurrence of a spike.  Both
stimulus reconstruction and the identification of triggering
features are affected by this issue. Some of the effects of
correlations between spikes have been treated in
\cite{gerstner,eguia,tiesinga,chacron}. To separate the influence
of previous spikes from the influence of the stimulus itself, we
propose to analyze only `isolated' spikes---spikes sufficiently
distant in time from previous spikes that they can be considered
to be statistically independent.  This procedure allows us to
recover exactly the relevant stimulus features.  However, it
introduces additional complications.  By requiring an extended
silence before a spike, we bias the prior statistical ensemble.
This bias will emerge in the white noise analysis, and we will
have to identify it explicitly.

We carry out this program for the simplest model neuron with
temporal dynamics: the leaky integrate-and-fire neuron.  We choose
this model because, unlike more complex models, we know exactly
which stimulus feature causes spiking, and can make a precise
comparison with the output of the analysis.

The methods presented and validated in the present work using the
integrate-and-fire model have been applied to the more
biologically interesting Hodgkin--Huxley neuron \cite{hh},
resulting in new insights into its computational properties.
Application of the same analytic techniques to real neural data is
currently underway.

\section{Characterizing neural response}

In this section we briefly review material that has appeared
elsewhere \cite{nips2000,hh,bill&robinprep}.

\subsection{The linear/nonlinear model}

Consider a neuron responding to a dynamic stimulus $I(t)$. While
here we will consider $I$ to be a scalar function, in general, it
may depend on any number of additional variables, such as
frequency, spatial position, or orientation. We can suppress these
dependencies without loss of generality. Even without additional
dependent variables, $I(t)$ is a high-dimensional stimulus, as the
occurrence of a spike at time $t_0$ generally depends on the
stimulus {\em history}, $I(t<t_0)$. The effective dimensionality
of this input is determined by the temporal extent of the relevant
stimulus, and the effective sampling rate, which is imposed by
biophysical limits. For the computation of the neuron to be
characterizable at all, the dimensionality of the stimulus
relevant to spiking must generally be much lower than this total
dimensionality.  The simplest approach to dimensionality reduction
is to approximate the relevant dimensions as a low-dimensional
linear subspace of $I(t)$.

If the neuron is sensitive only to a low dimensional linear
subspace, we can define a small set of signals $s_1 , s_2, \cdots
, s_K$ by filtering the stimulus,
\begin{equation}
s_\mu = \int_0^\infty d\tau f_\mu (\tau) I(t_0 - \tau) \, ,
\end{equation}
so that the probability of spiking depends only on these few
signals,
\begin{equation}
P[{\rm spike \, at \,} t_0 | I(t < t_0)] = P[{\rm spike \, at \,}
t_0] \, g(s_1, s_2, \cdots , s_K )\, . \label{e:Kprojs}
\end{equation}
Classic characterizations of neurons in the retina, for example
\cite{retinaISID}, typically assume that these neurons are
sensitive to a single linear dimension of the stimulus; the filter
$f_1$ then corresponds to the (spatiotemporal) receptive field,
and $g$ is proportional to the one-dimensional tuning curve.  A
number of other neurons, particularly early in sensory pathways,
have been similarly characterized, e.g.
\cite{theunissen,kim&rieke,liam}. Here we will concentrate
primarily on the problem of finding the filters $\{f_\mu\}$; once
they are found, $g$ can be obtained easily by direct sampling of
$P[{\rm spike\, at\, t_0}|s_1, s_2, \cdots , s_K]$, assuming that
the relevant dimensionality $K$ is small.

\subsection{Reverse correlation methods}

While one would like to determine the neural response in terms of
the probability of spiking given the stimulus, $P[{\rm spike \, at
\, } t_0 | I(t < t_0)]$, we follow \cite{bialek88} in considering
instead the distribution of signals conditional on the response,
$P[I(t < t_0)| {\rm spike \, at \, } t_0  ]$.  This quantity can
be extracted directly from the known stimulus and the resulting
spike train.  The two probabilities are related by Bayes' rule,
\begin{equation}
{{P[{\mathrm{spike} \, \mathrm{at} \,} t_0 | I(t <
t_0)]}\over{P[{\mathrm{spike} \, \mathrm{at} \, } t_0 ]}} =
{{P[I(t < t_0)| {\mathrm{spike} \, \mathrm{at} \,} t_0
]}\over{P[I(t < t_0)]}} \,. \label{e:Bayes}
\end{equation}
$P[\mathrm{spike} \, \mathrm{at} \, t_0]$ is the mean spike rate,
and $P[I(t < t_0)]$ is the prior distribution of stimuli.  If
$I(t)$ is Gaussian white noise, then this distribution is a
multidimensional Gaussian; furthermore, the distribution of any
filtered version of $I(t)$ will also be Gaussian.

The first moment of $P[I(t < t_0)| {\rm spike \, at \, } t_0 ]$ is
the spike-triggered average,
\begin{equation} \label{e:sta} \mathrm{STA}(\tau ) = \int [dI]
\,P[I(t < t_0)| {\rm spike \, at \,} t_0  ] I(t_0 -\tau) \, .
\end{equation}
We can also compute the covariance matrix of fluctuations around
this average,
\begin{equation}
\label{e:cov} C_{\rm spike} (\tau , \tau ') = \int [dI] \,P[I(t <
t_0)| {\rm spike \, at \,} t_0  ] I(t_0 -\tau) I(t_0 -\tau') -
\mathrm{STA}(\tau ) \mathrm{STA}(\tau ') .
\end{equation}
In the same way that we compare the spike-triggered average to
some constant average level of the signal in the whole experiment,
an important advance of \cite{naama-info-method} was to compare
the covariance matrix $C_{\rm spike}$ with the covariance of the
signal averaged over the whole experiment,
\begin{equation}
C_{\rm prior} (\tau , \tau ')= \int [ds] \,P[s(t < t_0)]  s(t_0
-\tau) s(t_0 -\tau') \, .
\end{equation}

If the probability of spiking depends on only a few relevant
features, the {\em change} in the covariance matrix, $\Delta C =
C_{\rm spike} - C_{\rm prior}$, has a correspondingly small number
of outstanding eigenvalues. Precisely, if $P[{\rm spike}|{\rm
stimulus}]$ depends on $K$ linear projections of the stimulus as
in Eq. (\ref{e:Kprojs}), and if the inputs $I(t)$ are chosen from
a Gaussian distribution, then the rank of the matrix $\Delta C$ is
exactly $K$.  Further, the eigenmodes associated with nonzero
eigenvalues span the relevant subspace.  A positive eigenvalue
indicates a direction in stimulus space along which the variance
is increased relative to the prior, while a negative eigenvalue
corresponds to reduced variance.

\section{Leaky integrate-and-fire}
\label{s:ifmodel}

The leaky integrate-and-fire neuron is perhaps the most commonly
used approximation to real neurons.  Its dynamics can be written:
\begin{equation}\label{e:ifmodel}
 C{{dV}\over{dt}} = I(t) - {V\over R} - C V_c \sum _i \delta(t-t_i) \, ,
\end{equation}
where $C$ is a capacitance, $R$ is a resistance and $V_c$ is the
voltage threshold for spiking.  The first two terms on the right
hand side model an RC circuit; the capacitor integrates the input
current as a potential $V$ while the resistor dissipates the
stored charge (hence `leaky').  The third term implements spiking:
spikes are defined as the set of times $\{t_i\}$ such that
$V(t_i)=V_c$.  When the potential reaches $V_c$, a restoring
current is instantaneously injected to bring the stored charge
back to zero, resetting the system.  This formulation emphasizes
the similarity between leaky integrate-and-fire and the more
biophysically realistic Hodgkin--Huxley equation for current flow
through a patch of neural membrane; in the Hodgkin--Huxley system,
however, the nonlinear spike-generating term is replaced with
ionic currents of a more complex (and less singular) form. Hence
the simple properties of the leaky integrate-and-fire model---a
single degree of freedom $V$, instantaneous spiking, total
resetting of the system following a spike, and linearity of the
equation of motion away from spikes---are only rough
approximations to the properties of realistic neurons.

Integrating Eq. (\ref{e:ifmodel}) away from spikes, we obtain
\begin{equation} \label{e:ifsol}
CV(t)=\int _{-\infty} ^t d\tau \exp \left( {\tau - t} \over {RC}
\right) I(\tau) \, ,
\end{equation}
assuming initialization of the system in the distant past at zero
potential.  The right hand side can be rewritten as the
convolution of $I(t)$ with the causal exponential kernel
\begin{equation} \label{e:ifkernel}
f_\infty(\tau)=\left\{ \begin{array}{ll}
 0               & \mbox{if $\tau \leq 0$}\\
 \exp(-\tau /RC) & \mbox{if $0 < \tau$,}
 \end{array}
 \right.
\end{equation}
so the condition for a spike at time $t_1$ is that this
convolution reach threshold:
\begin{equation} \label{e:ifcond}
CV_c=\int _{-\infty} ^\infty d\tau f_\infty(\tau) I(t_1-\tau) \, .
\end{equation}
The interaction between spikes complicates this picture somewhat.
If the system spiked last at time $t_0 < t_1$, we must replace the
lower limit of integration in Eq. (\ref{e:ifsol}) with $t_0$, as
no current injected before $t_0$ can contribute to the accumulated
potential. However, if we wish always to evaluate stimulus
projections by integrating over the entire current history $I(t)$
as in Eq. (\ref{e:ifcond}), then we must replace $f_\infty$ with
an entire family of kernels dependent on the time to the previous
spike, $\Delta t=t_1-t_0$:
\begin{equation} \label{e:iffamily}
f_{\Delta t}(\tau)=\left\{ \begin{array}{lll}
 0              & \mbox{if $\tau\leq 0$}\\
 \exp(-\tau/RC) & \mbox{if $0 < \tau < \Delta t$}\\
 0              & \mbox{if $\Delta t \leq \tau$.}
 \end{array}
 \right.
\end{equation}
This equation completes an exact model of the leaky
integrate-and-fire neuron.

\section{Response to white noise}
\label{s:ifwhitenoise}

We now consider the statistical behavior of the integrate-and-fire
neuron when driven with Gaussian white noise current of standard
deviation $\sigma$,
\begin{eqnarray} \label{e:ifinput}
 \langle I(t) \rangle &=& 0, \nonumber \\
 \langle I(t)I(t') \rangle &=& \sigma ^2 \delta (t-t').
\end{eqnarray}
For the following analysis, we take $R=10$ k$\Omega$, $C=1$
$\mu$F, $V_c=10$ mV and $\sigma=\sqrt{200}$ $\mu$A, and use Euler
integration with a time step of 0.05 msec.  These values were
chosen to mimic biophysically plausible parameters for a 1
$\mathrm{cm}^2$ patch of membrane, but none of our results depend
qualitatively on this choice.

\begin{figure}[hpt]
\begin{center}
\includegraphics[scale=.8]{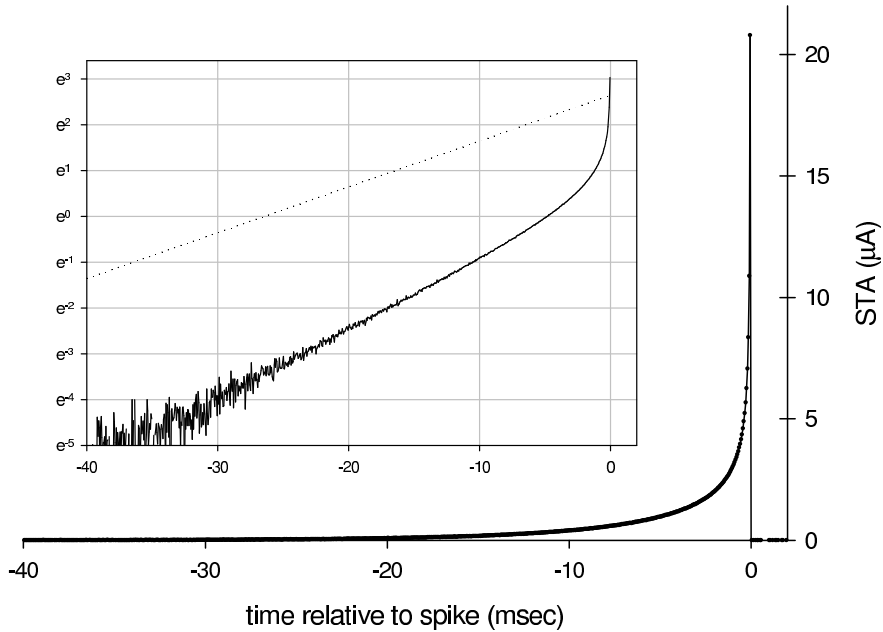}
\end{center}
\caption{Spike-triggered average of a leaky integrate-and-fire
neuron; shown inset with a logarithmic $I$-axis and the pure
exponential $\sigma f_\infty$ (dotted line). } \label{f:rawsta}
\end{figure}

\subsection{Unqualified reverse correlation}

At first glance, it would appear that the neuron is perfectly
described by a reduced model in one dimension: the current
filtered by the exponential kernel of Eq. (\ref{e:ifkernel}). If
so, the spike-triggered average under white noise stimulation
should be proportional to this filter.\footnote{In possible linear
combination with a derivative-like filter, arising from the
additional constraint that the threshold must be crossed from
below.  This point will be addressed in more detail later.}
However, as is evident in Fig. \ref{f:rawsta}, the STA is not
exponential, nor does it asymptote to the appropriate decay rate
$1/RC$.  One reason is that, in the presence of previous spikes,
the integrate-and-fire neuron is {\em not} low-dimensional. Recall
that the relative timing $\Delta t$ of the previous spike selects
a filter $f_{\Delta t}$ (Eq. (\ref{e:iffamily})); these filters
are truncated versions of the original exponential $f_\infty$. The
STA will therefore include contributions from each of these
filters, weighted by their probability of occurrence (as
determined by the interspike interval distribution).\footnote{This
is not a prescription for recovering $f_\infty$ from the STA: the
STA is further influenced by other effects, including the average
stimulus histories leading up to {\em previous} spikes.}  It is
readily shown that the orthogonal component of each filter
$f_{\Delta t}$ is a $\delta$-function at $-\Delta t$; so the set
of relevant filters spans the entire stimulus space.

\begin{figure}[hpt]
\begin{center}
\includegraphics[scale=0.64]{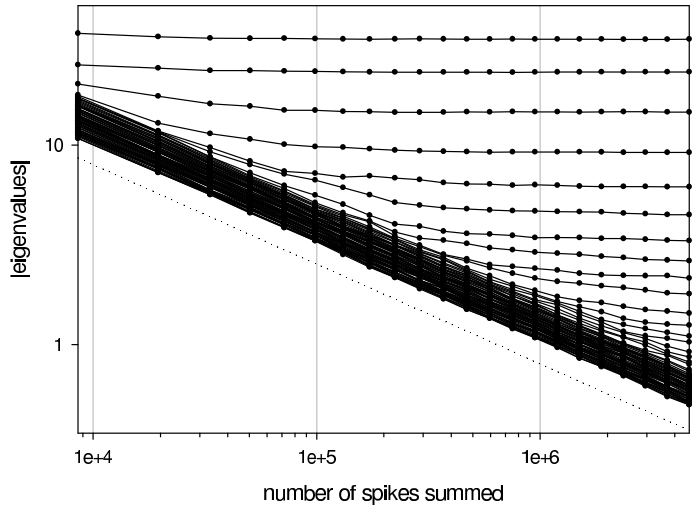}
\end{center}
\caption{Eigenvalue magnitudes of the spike-triggered covariance
difference matrix $\Delta C$ as a function of the number of spikes
used to accumulate the matrix.  An arbitrary number of stable
modes emerge, given enough data.  The stable modes all have
negative eigenvalues, i.e. spikes are associated with reduced
stimulus variance along these dimensions.  The dotted line, shown
for reference, is proportional to $1/\sqrt{N_{\rm spikes}}$. }
\label{f:covall}
\end{figure}

\begin{figure}[hpt]
\begin{center}
\includegraphics[scale=0.8]{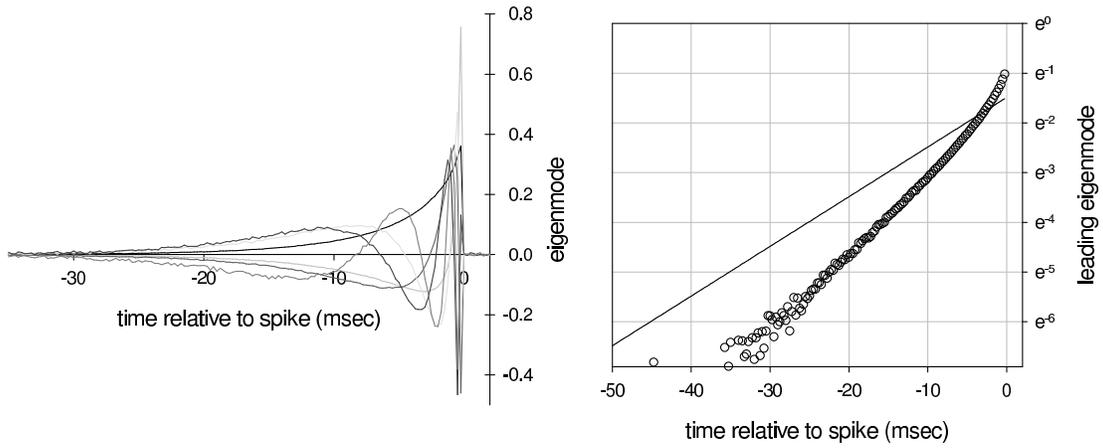}
\end{center}
\caption{Left: the leading six modes of the unqualified
spike-triggered change in covariance $\Delta C$.  Right: only the
first mode is monophasic; here it is plotted on a logarithmic
$y$-axis (circles) with $f_\infty$ for reference (solid line).}
\label{f:rawmodes}
\end{figure}

Covariance analysis reveals this high dimensionality very clearly.
In Fig. \ref{f:covall} we show the spectrum of eigenvalue
magnitudes of $\Delta C$ as a function of the number of spikes
summed to accumulate the matrix; with decreasing noise, an
arbitrary number of significant eigenmodes emerge.  As shown in
Fig. \ref{f:rawmodes}, none of these eigenmodes correspond to an
exponential filter.  The first mode most closely resembles an
exponential, but as shown in the right half of Fig.
\ref{f:rawmodes}, it is more peaked near $t=0$, and its time
constant is very different from $RC$.

\subsection{Isolated spikes}

\begin{figure}[hpt]
\begin{center}
\includegraphics[scale=0.65]{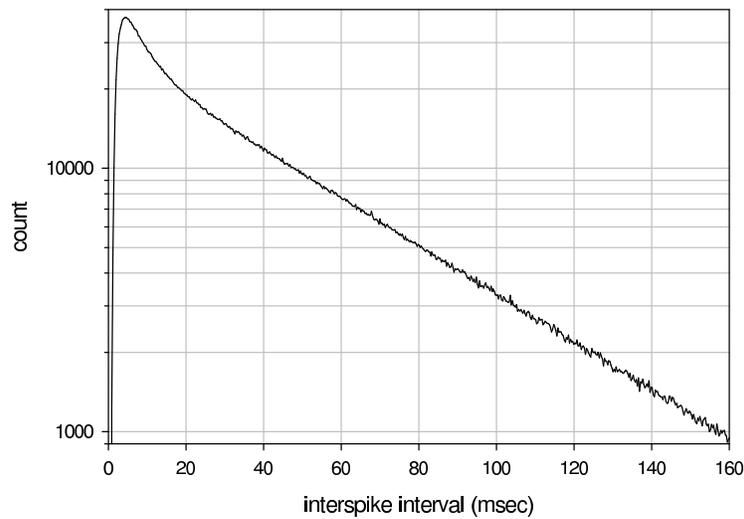}
\end{center}
\caption{Interspike interval histogram for a leaky
integrate-and-fire neuron.  $5\times 10^6$ spikes were accumulated
in $0.27$ msec bins.} \label{f:isid}
\end{figure}

It is clear that making progress requires first considering the
causes of a single spike independently of the spiking history.
Therefore we will include in our analysis only spikes which follow
previous spikes after a sufficiently long interval.  The
appropriate interval can be estimated from the interspike interval
distribution, Fig. \ref{f:isid}.

As in many real neurons, the interval distribution has three main
features: a `refractory period' during which spikes are strongly
suppressed, a peak at a preferred interval, and an exponential
tail. Intervals in the exponential tail follow Poisson statistics;
for intervals in this range, we can take spikes to be
statistically independent.  With reference to Fig. \ref{f:isid},
we set our isolated spike criterion---very conservatively---as 75
msec of previous `silence'.

\begin{figure}[hpt]
\begin{center}
\includegraphics[scale=0.7]{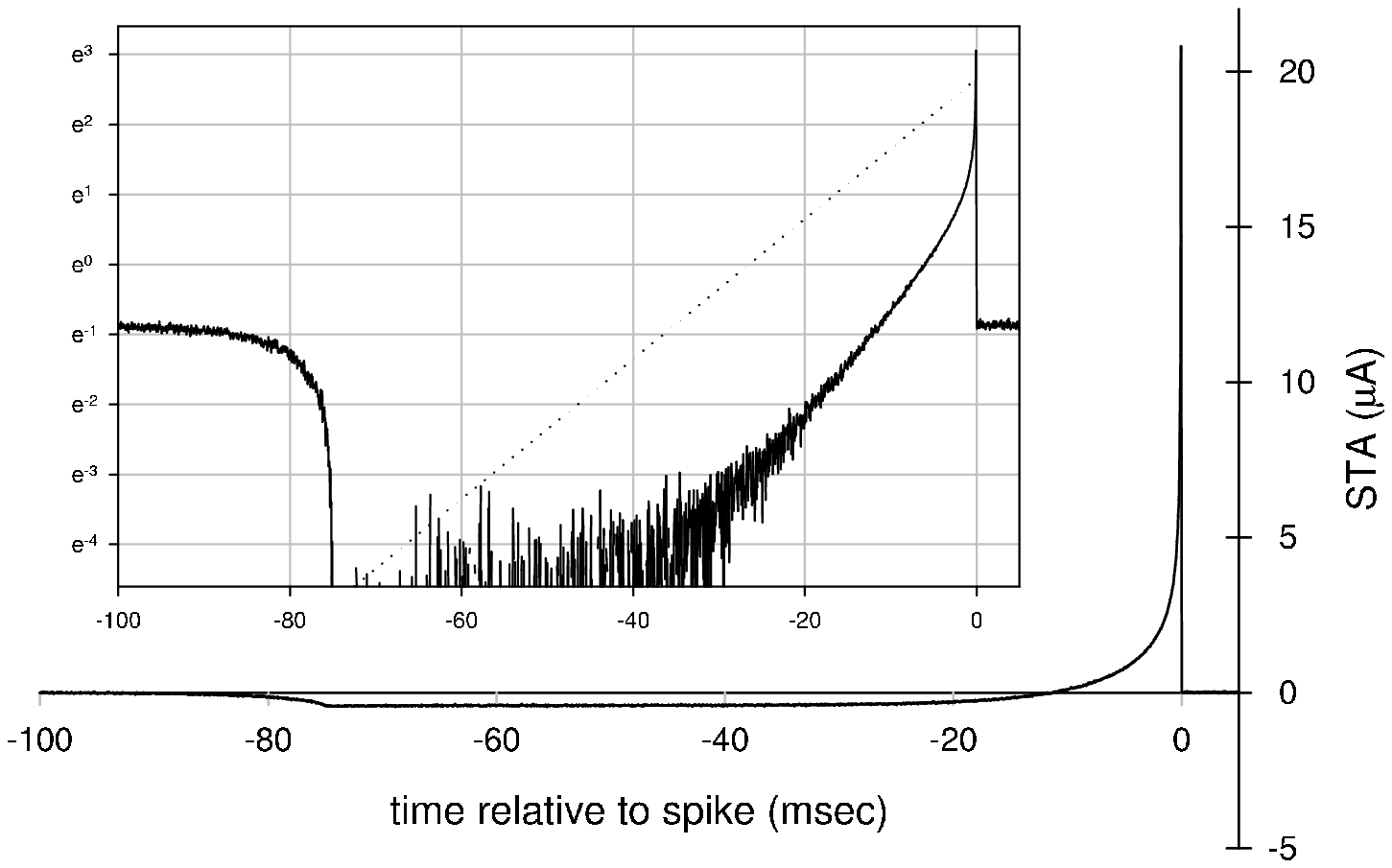}
\end{center}
\caption{Triggered average stimulus for isolated spikes. Isolated
spikes are defined as spikes which are preceded by at least 75
msec of silence.  The inset shows the triggered average with the
constant silence bias subtracted on a logarithmic $I$-axis, again
with the pure exponential $\sigma f_\infty$ drawn as a dotted line
for reference.} \label{f:isosta}
\end{figure}

As shown in Fig. \ref{f:isosta}, even with this constraint, we do
not recover a spike-triggered average matching $f_\infty$. We
should note that this is no longer, strictly speaking, a
spike-triggered average, but an {\em event}-triggered average,
where the event includes the preceding silence.  Thus we can
expect the silence itself to contribute to any average we
construct.  To first order, during silence there is a constant
negative bias, since certain positive currents which would have
caused a spike are excluded from the average.  Similarly, we
expect that the covariance of the current will be altered from
that of the Gaussian prior during extended silences. Because
extended silence is not locked to any particular moment in time,
the covariance matrix is stationary during silence.

\begin{figure}[hpt]
\begin{center}
\includegraphics[scale=0.6]{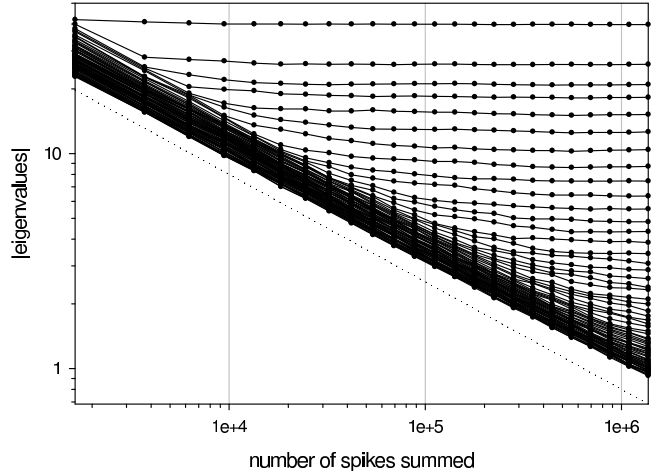}
\end{center}
\caption{Eigenvalue magnitudes of the isolated spike-triggered
change in covariance $\Delta C_{\rm iso}$ as a function of the
number of isolated spikes used to accumulate the matrix.  The
stable modes again have negative eigenvalues.} \label{f:isomodes}
\end{figure}

We construct the covariance matrix as before: $\Delta C_{\rm
iso}=C_{\rm isolated\, spike}-C_{\rm prior}$.  An entire series of
modes once again appears, Fig. \ref{f:isomodes}.  This time, they
arise because silence is, by definition, translationally
invariant, resulting in a Fourier-like spectrum.  Modes associated
with the spike, on the other hand, are locked to a particular
time, and should therefore not have Fourier-like modes.  In
particular, these spiking modes should have temporal support only
in the time immediately before the spike. This implies that two
types of modes---silence- and spike-associated---should emerge
independently in the covariance analysis, and should exhibit a
clear difference in their domain of support.

\begin{figure}[hpt]
\begin{center}
\includegraphics[scale=0.65]{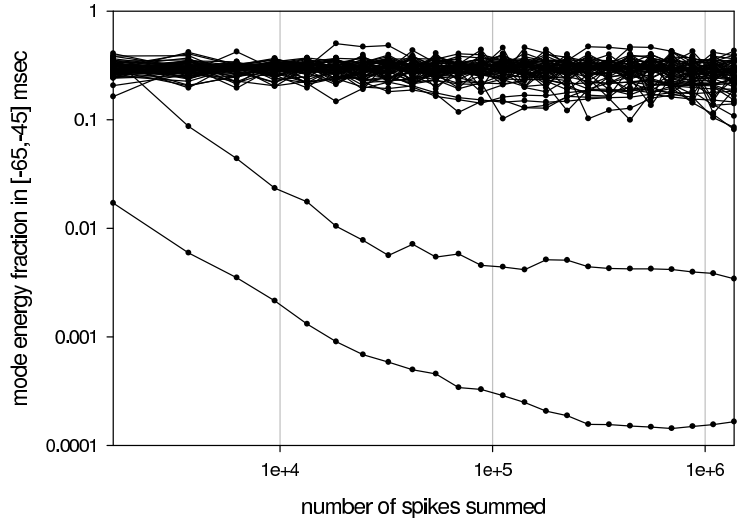}
\end{center}
\caption{Fraction of the energy in each mode of $\Delta C_{\rm
iso}$ (modes are defined over $t\in [-65,0]$ msec) in the silent
interval $t\in [-65,-45]$ msec, as a function of the number of
isolated spikes used to accumulate $\Delta C_{\rm iso}$.  The
eigenvalue for a mode with zero energy in this interval would
decrease like $1/N_{\rm spikes}$, as the only contribution would
come from noise. Here, two modes emerge with very small energy in
this interval, though they stabilize at small nonzero values.
These modes are localized to the time immediately before the
spike, but as they have exponential support, their energy in the
silence does not fall all the way to zero.} \label{f:powerfrac}
\end{figure}

In Fig. \ref{f:powerfrac}, we consider the fraction of the energy
of each mode over the interval $t \in [-65,-45]$ msec.  At $-45$
msec, the exponential filter $f_\infty$ has decayed to 1\% of its
peak value, so we are well away from the spike at $t=0$.  The 75
msec silence criterion also puts the lower boundary of this
interval well inside the stationary silence (at least 10 msec
after the previous spike). As we see, exactly two modes emerge as
local, with their energy fraction decaying with the noise in the
matrix until reaching a stable value well below the rest of the
modes.

\begin{figure}[hpt]
\begin{center}
\includegraphics[scale=0.7]{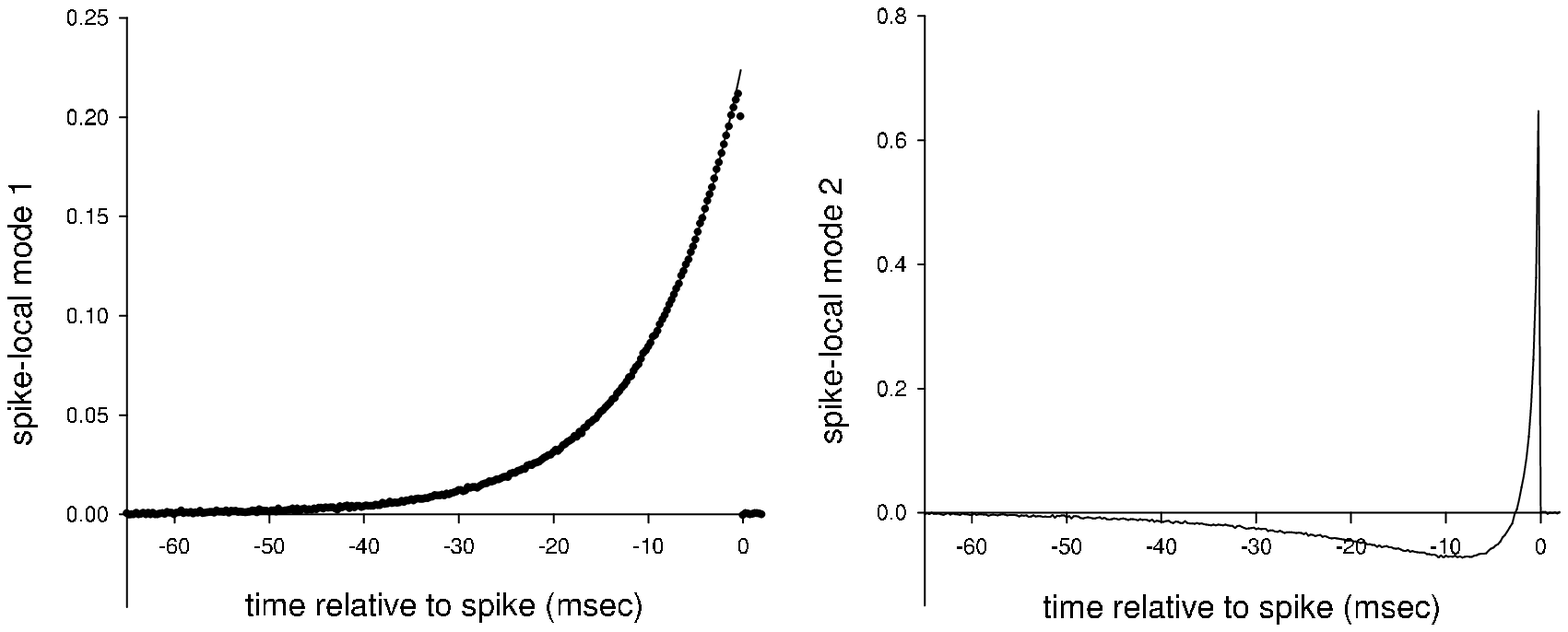}
\end{center}
\caption{The two spike-associated modes of $\Delta C_{\rm iso}$.
The first (left) is the recovered exponential filter of the leaky
integrate-and-fire spiking condition; the measured mode is marked
with dots, and the line is $f_\infty$ (normalized). The second
mode (right) is the `first crossing' constraint.}
\label{f:spikemodes}
\end{figure}

The first localized mode, shown in Fig. \ref{f:spikemodes}, is the
long-awaited exponential filter $f_\infty$.  Using our modified
second-order reverse correlation analysis on `experimental data'
we have thus finally been able to recover the stimulus feature to
which the leaky integrate-and-fire neuron is sensitive.

\subsection{Constraints}

The second mode is a consequence of an implicit constraint in the
spiking model.  Na\"ively, we know that at the moment when
$V=V_c$, $\dot V>0$; that is, the threshold must be crossed from
below.  Hence
\begin{equation} \label{e:ifdcond}
\int _{-\infty} ^{\infty} d\tau I(t-\tau) {d\over{d\tau}}
f_\infty(\tau) < 0,
\end{equation}
making the neuron also sensitive to the time derivative of
$f_\infty$.  The integrate-and-fire neuron is peculiar in that
$f_\infty$ and $df_\infty/d\tau$ are almost linearly dependent:
\begin{equation} \label{e:ifdf0}
{d\over{d\tau}}f_\infty (\tau)=\left\{ \begin{array}{ll}
 \delta (\tau)             & \mbox{if $\tau \leq 0$.}\\
 -(RC)^{-1}\exp(-\tau /RC) & \mbox{if $0 < \tau$.}
 \end{array}
 \right.
\end{equation}
The orthogonal part of this filter is therefore only $\delta
(\tau)$, and we can rewrite this extra condition more simply as
$I(0)>V_c/R$.

However, we see that although the second spiking mode is indeed
very sharply peaked at $\tau =0$, it is not a $\delta$-function.
While the above argument is correct in constraining the possible
values of $I(0)$, we have also required no threshold crossings for
an extended time prior to the spike. Hence certain trajectories
$I(t)$ which satisfy the two basic conditions ($V(0)=V_c$ and
$\dot V(0)>0$) are still disallowed, because they would have led
to a spike earlier.  The second mode is therefore an extended
version of the derivative condition.  Although it appears to have
very extended support, this is due to orthogonalization with
respect to the first mode, $f_\infty$. The true constraint is
positive definite and highly peaked, with virtually all of its
energy concentrated in the $\sim 5$ msec interval prior to the
spike.

\begin{figure}[hpt]
\begin{center}
\includegraphics[scale=0.65]{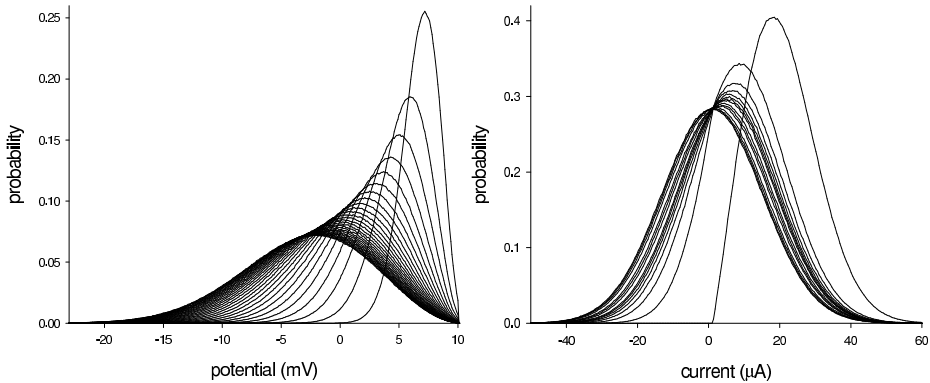}
\end{center}
\caption{The time-dependent distributions of $V(t)$ (left) and
$I(t)$ (right) leading up to an isolated spike at $t=0$.  In both
cases, the distributions with the highest means correspond to the
final time slice (they are thus also the most constrained).  For
$V$, the times shown are $\{-15,-14.5,-14,\,\cdots\, ,
-1.5,-1,-0.5\}$ msec.  For $I$, the times are $\{-5.05, -4.05,
\,\cdots\, , -1.05\}$, $\{-0.8, -0.6, -0.4\}$, and $\{-0.35, -0.3,
\,\cdots\, , -0.05\}$ msec. The distribution of $V$ is virtually
indistinguishable from its silence steady-state by $-15$ msec;
convergence to the silence steady-state is much faster for $I$,
occurring by $-5$ msec.} \label{f:distribs}
\end{figure}

We can more easily understand this constraint by considering the
time-dependent distribution of $V(t)$ leading up to an isolated
spike, shown in the left panel of Fig. \ref{f:distribs}. The
process $V(t)$ leading up to a spike is a centrally biased random
walk with an absorbing barrier at $V_c$ and a capacitatively
determined correlation time.  If we trace the distribution of $V$
backward in time from the spike at $t=0$, we see a constrained
diffusive process from a $\delta$-function at the spike time to
the steady-state silence distribution.  The constraint on $V$
enforces a corresponding constraint on $I$, whose time-dependent
distribution is shown in the right panel of Fig. \ref{f:distribs}.
Notice that the $I$ distribution is only noticeably distorted from
its silence steady-state at times very near the spike.

\subsection{Silence modes}

\begin{figure}[hpt]
\begin{center}
\includegraphics[scale=0.65]{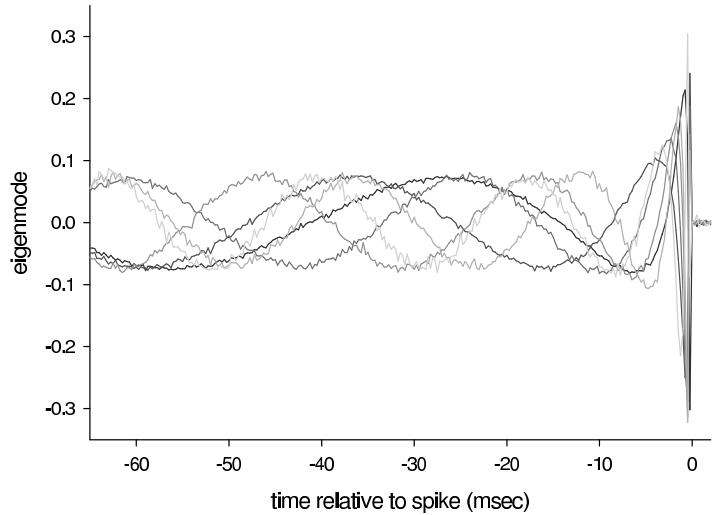}
\end{center}
\caption{The leading six modes of $\Delta C_{\rm iso}$ associated
with silence (i.e. with nonvanishing support over $t \in
[-65,-45]$ msec).  Although they are Fourier-like in the silence,
they chirp leading up to the spike.} \label{f:silencemodes}
\end{figure}

Finally, let us return to the spectrum of silence-associated modes
in Fig. \ref{f:powerfrac}.  Several of these are shown overlaid in
Fig. \ref{f:silencemodes}. It is immediately clear that they have
the expected Fourier-like structure over the period of silence.
Their behavior near the spike is less obvious, however: each mode
appears to execute an FM chirp, culminating in a sharp feature at
the spike time itself.  Clearly, while it is true by construction
that spike-associated covariance modes have no support in periods
of extended silence, the converse is not true.  This structure is
due to the essential {\em non}-stationarity of silence near
spikes: at some point, the silence must come to an abrupt end.

It is clear that in order to implement our procedure for
separating silence from spikes, we could not have taken the
approach of looking for modes with significant energy near the
spike.  All of the silence modes have interesting structure near
the spike, and could easily be mistaken for additional
spike-generating features.  Like the spike-associated modes, these
silence modes have negative eigenvalues, indicating reduced
stimulus variance in certain dimensions.  Yet they do not cause
spiking, nor do they inhibit it; they are in no meaningful sense
`suppressive' modes. They simply show us the second-order
statistics of silences, which emerge from the enforced {\em
absence} of spikes. This underscores the necessity of considering
a sufficiently extended silence before isolated spikes for the
translationally invariant part of the silence modes to emerge,
distinguishing them from spiking modes.

This structure also highlights the reason why we could not, as one
might expect, simply subtract from $C_{\rm isolated\, spike}$ a
prior obtained from extended silences instead of $C_{\rm prior}$
to obtain a simple, low-dimensional $\Delta C_{\rm iso}$.  While
such a subtraction results in a local matrix around the spike, it
cannot take into account the non-stationary part of the silence,
and leaves the spike-related and non-stationary silence-related
aspects of the covariance near the spike mixed.  This again
results in a high-dimensional matrix with incorrect (mixed) modes.

\section{Adaptation}

It has previously been observed that the STAs of some neurons
depend on the stimuli used to probe their response
\cite{theunissen,robnetwork}.  White noise stimulus was originally
introduced to circumvent this difficulty by sampling stimulus
space in an unbiased way; however, one must still choose a noise
variance. While one would hope that this does not affect the STA,
it has been shown that, at least in some sense, even the
integrate-and-fire neuron `adapts' to the stimulus variance
\cite{rudd,liam}, showing both a change in the overall STA and in
its measured nonlinear decision function. The reason for this form
of adaptation is simple: the statistics of spike intervals depend
on the stimulus variance, as the time to reach threshold depends
on the variance.  By construction, however, neither the
spike-triggering feature $f_\infty$ nor the decision function $g$
(here simply the threshold) of the integrate-and-fire neuron
actually depends on the stimulus statistics. As we have discussed,
the unqualified STA is a linear combination of the filters
$f_{\Delta t}$ conditioned on the time to the last spike; the
contribution of each filter to the overall STA thus depends on the
probability $P(\Delta t)$ of the interspike interval $\Delta t$.
As different stimulus ensembles produce different interspike
interval distributions, these coefficients are functions of the
stimulus ensemble. Therefore even in the absence of adaptation in
the linear filters $f$, the overall STA will be
stimulus-dependent. Because sampling the input/output relation
normally involves convolving the stimulus with the STA, the
measured nonlinearity will also appear to be stimulus-dependent,
despite the fact that the neuron's nonlinear decision function $g$
(based on convolution of the stimulus with the fixed filters $f$)
is also fixed.  Hence this form of `adaptation' does not reflect
any changing property or internal state of the neuron in response
to the stimulus statistics.  Rather, it arises from the
nonlinearity inherent in spiking itself.

This emphasizes the significance of our methods: we have
introduced a way to extract from data the intrinsic, unique
spike-triggering feature, removing the effects of the
stimulus-dependent interspike interval statistics.  These
statistics are produced by the model when driven by a particular
stimulation regime; they are not part of the model itself. If a
neuron does in fact exhibit `true' adaptation in the sense that
the functions $f$ and $g$ are variance-dependent, then the
isolated spike analysis presented here will reveal this. Such
variance dependence can come about as a result of explicit gain
control mechanisms, e.g. \cite{shapley}, or because nonlinearities
in the subthreshold neural response select non-constant stimulus
features, an effect which can play an important role in more
complex models such as the Hodgkin--Huxley neuron \cite{hh}.

\section{Discussion}

While the interspike interval statistics are not an intrinsic
model property, they do depend on an aspect of the complete
characterization of the neuron that we have explicitly neglected
here: the interspike interaction.  There are two senses in which
addressing only isolated spikes appears to be a limitation of our
method.  One is practical: in most situations, this will
substantially reduce the number of spikes considered in the
analysis.  Further, much of the work on white noise analysis to
date has focussed on the problem of stimulus reconstruction, and
our method does not provide a recipe for decoding non-isolated
spikes.  With respect to the second point, there are good reasons
for considering isolated spikes first.  Evidence from several
systems suggests that isolated spikes contain more information
about the stimulus than the spikes that follow them
\cite{retinaISID,rob,ras}.  This is clear from first principles,
as non-isolated spikes are triggered only partly by the stimulus,
and partly by the timing of the previous spike. Understanding the
complex interactions between spikes and the stimulus to create
spike patterns is obviously an important next step in
understanding the neural code, but the logical starting point is
to establish the causes of isolated spikes.  If we attempt
stimulus reconstruction using reverse correlations averaged over
all spikes, then the reconstruction will be incorrect; similarly,
predicting spike timing without knowledge of the recent spiking
history is generally impossible. A step along the path of treating
spike interactions explicitly has recently been made in the
modelling of retinal ganglion cell spike trains \cite{everyspike}.

The separation of spikes into isolated and non-isolated also
allows a more efficient use of the data---ensemble averages will
not be complicated by the inclusion of effects from interspike
interaction.  Thus although we are using less data, it is probably
the case that we are using the data more effectively, by summing
stronger covariance signals from a simpler, lower-dimensional
space.  While in a simulation the amount of data can be unbounded,
and we have used up to millions of spikes to demonstrate our
points here, we note that the significant effects all emerge very
clearly after $\sim 10^4$ spikes.  This can be seen from the
evolution of the eigenvalues and silence energy fractions with
increasing spike count (Figs. \ref{f:covall}, \ref{f:isomodes} and
\ref{f:powerfrac}).  This number is in a range often accessible to
the experimenter.

It should be noted that the need to consider the effect of silence
arises even in the standard analysis of non-isolated spikes.
Because all spiking neurons have a refractory period, there is
always an implicit silence or isolation constraint, albeit
sometimes of only a few milliseconds.  The second spike-associated
mode identified here (Fig. \ref{f:spikemodes}) is localized
precisely in the disallowed interval, and in fact emerges almost
identically in the unqualified covariance analysis.  Its
contribution to the highly peaked STA is very significant, since
both the spike-conditional variance and mean along this stimulus
dimension are substantially altered from the prior. Recall that
this mode is related to a derivative condition expressing upward
threshold crossing, but it is extended, as there is always a
finite period of silence prior to a spike.  An analogous mode
should arise for all spiking neurons, though its shape will depend
strongly on the shape of the spike-generating filter or filters.
Reduced stimulus variance in the direction of this derivative-like
feature should not be thought of as part of the {\em cause} of a
spike; it is a necessary concomitant to spike generation
conditioned on the first filter alone. This inevitable
derivative-like condition is relevant to the inverse problem of
stimulus reconstruction from a spike train, but not to the forward
problem of causally predicting spike times from the stimulus.
Notice that this implies that, in general, convolving the stimulus
with the STA---either unqualified or for isolated spikes only---is
not an ideal way to predict spike times.

In summary, we have shown that standard white noise analysis is
unable to recover the known filter properties of the leaky
integrate-and-fire neuron.  We show that this is due to the
influence of interactions between spikes, and we demonstrate that
by removing this influence, requiring spikes to be `isolated', we
are able to recover the computation performed by the neuron.
However, in so doing, we introduce a complication: we must address
not only the isolated spikes, but the silences that precede them.
Silences produce complex structure in the covariance analysis,
which is nonetheless statistically independent and clearly
distinguishable from structure tied to the spikes themselves.  In
a sense, the distinction between spike-related and silence-related
features is artificial, for silence is simply the complement to
spiking: silence is structured by the absence of features which
would produce a spike. The distinction is useful, however, in
allowing us to use the isolated spikes to identify the
spike-generating feature or features unambiguously.

We have shown that failing to consider spikes in isolation leads
to distortion in the filters obtained through standard white noise
reverse correlation.  Our conclusions are relevant not just for
the simple case of the leaky integrate-and-fire neuron, but for
any system in which the presence of a spike has an influence on
the generation of subsequent spikes.  The new isolated spike
method presented should improve the accuracy of white noise
analysis of neural spike trains with interspike interactions. In
some cases, it may significantly change our picture of what
stimulus features a neuron is sensitive to.

\vspace{1cm}

\begin{quote}
\it \small We would like to thank William Bialek for his advice,
discussion, and encouragement to study this problem.
\end{quote}

\vfill\newpage

{\small
\bibliographystyle{unsrt}
\bibliography{2003_Aguera_Fairhall_IntAndFire}
}

\vfill

\end{document}